\documentclass[12pt]{article}
\usepackage{epsfig}

\def\lqcd{\Lambda_{\rm QCD}}
\def\qsh{\hat q^2}
\def\qcut{q_0}

\def\gtrsim{\mathrel{\mathpalette\vereq>}}
\def\vereq#1#2{\lower3pt\vbox{\baselineskip1pt\lineskip1pt
    \ialign{\\$#1\hfill##\hfil\\$\crcr#2\crcr\sim\crcr}}}

\newcommand\pubnumber{UTPT-01-03\\UCSD/PTH 01-02\\ LBNL--47132}
\newcommand\pubdate{\today}
\newcommand\hepnumber{hep-ph/0101328}

\def\support{\footnote{${}^*$Talk given by M.L.}}

\def\Title#1{\begin{center} {\Large\bf #1 } \end{center}}
\def\Author#1{\begin{center}{ \sc #1} \end{center}}
\def\Address#1{\begin{center}{ \it #1} \end{center}}

\newcommand\pubblock{\rightline{\begin{tabular}{l} \pubnumber\\
         \pubdate\\ \hepnumber \end{tabular}}}
\newenvironment{Abstract}{\begin{quotation}  }{\end{quotation}}
\newenvironment{Presented}{\begin{quotation} \begin{center} 
             Presented at the\end{center}
      \begin{center}\begin{large}}{\end{large}\end{center} \end{quotation}}

\makeatletter
\def\section{\@startsection{section}{0}{\z@}{5.5ex plus .5ex minus
 1.5ex}{2.3ex plus .2ex}{\large\bf}}
\def\subsection{\@startsection{subsection}{1}{\z@}{3.5ex plus .5ex minus
 1.5ex}{1.3ex plus .2ex}{\normalsize\bf}}
\def\subsubsection{\@startsection{subsubsection}{2}{\z@}{-3.5ex plus
-1ex minus  -.2ex}{2.3ex plus .2ex}{\normalsize\sl}}

\renewcommand{\@makecaption}[2]{%
   \vskip 10pt
   \setbox\@tempboxa\hbox{\small #1: #2}
   \ifdim \wd\@tempboxa >\hsize     
       \small #1: #2\par          
     \else                        
       \hbox to\hsize{\hfil\box\@tempboxa\hfil}
   \fi}

 \def\citenum#1{{\def\@cite##1##2{##1}\cite{#1}}}
 
\newcount\@tempcntc
\def\@citex[#1]#2{\if@filesw\immediate\write\@auxout{\string\citation{#2}}\fi
  \@tempcnta\z@\@tempcntb\m@ne\def\@citea{}\@cite{\@for\@citeb:=#2\do
    {\@ifundefined
       {b@\@citeb}{\@citeo\@tempcntb\m@ne\@citea\def\@citea{,}{\bf ?}\@warning
       {Citation `\@citeb' on page \thepage \space undefined}}%
    {\setbox\z@\hbox{\global\@tempcntc0\csname b@\@citeb\endcsname\relax}%
     \ifnum\@tempcntc=\z@ \@citeo\@tempcntb\m@ne
       \@citea\def\@citea{,}\hbox{\csname b@\@citeb\endcsname}%
     \else
      \advance\@tempcntb\@ne
      \ifnum\@tempcntb=\@tempcntc
      \else\advance\@tempcntb\m@ne\@citeo
      \@tempcnta\@tempcntc\@tempcntb\@tempcntc\fi\fi}}\@citeo}{#1}}
\def\@citeo{\ifnum\@tempcnta>\@tempcntb\else\@citea\def\@citea{,}%
  \ifnum\@tempcnta=\@tempcntb\the\@tempcnta\else
  {\advance\@tempcnta\@ne\ifnum\@tempcnta=\@tempcntb \else\def\@citea{--}\fi
    \advance\@tempcnta\m@ne\the\@tempcnta\@citea\the\@tempcntb}\fi\fi}
\makeatother

\def\beq{\begin{equation}}
\def\eeq#1{\label{#1}\end{equation}}
\def\eeqn{\end{equation}}

\newenvironment{Eqnarray}%
   {\arraycolsep 0.14em\begin{eqnarray}}{\end{eqnarray}}
\def\beqa{\begin{Eqnarray}}
\def\eeqa#1{\label{#1}\end{Eqnarray}}
\def\eeqan{\end{Eqnarray}}

\let\bar=\overbar

\def\Dslash{\not{\hbox{\kern-4pt $D$}}}
\def\dslash{\not{\hbox{\kern-2pt $\del$}}}

\def\msb{{\bar{\ssstyle M \kern -1pt S}}}

\def\lsim{\mathrel{\raise.3ex\hbox{$<$\kern-.75em\lower1ex\hbox{$\sim$}}}}
\def\gsim{\mathrel{\raise.3ex\hbox{$>$\kern-.75em\lower1ex\hbox{$\sim$}}}}

\begin{document}
\begin{titlepage}
\pubblock

\vfill
\def\thefootnote{\fnsymbol{footnote}}
\Title{Determining $|V_{ub}|$ from the $\bar B\to X_u\ell\bar\nu$ 
dilepton invariant mass spectrum$^{\,*}$\support}
\vfill
\Author{Christian W.\ Bauer,$^1$ 
  Zoltan Ligeti,$^2$ and Michael Luke$^{3}$}
\Address{ \vspace*{4pt}
  $^1$Department of Physics, University of California, San Diego,\\
9500 Gilman Drive, La Jolla CA  USA 92093 \\[3pt]
  $^2$Theoretical Physics Group \\
    Ernest Orlando Lawrence Berkeley National Laboratory \\
    University of California, Berkeley, CA USA 94720 
\\ [3pt]$^3$Department of Physics, University of Toronto, \\
    60 St.\ George Street, Toronto, Ontario, Canada M5S 1A7 }
\vfill
\begin{Abstract}
The invariant mass spectrum of the lepton pair in
inclusive semileptonic $\bar B\to X_u \ell\bar\nu$ decay yields a model
independent determination of $|V_{ub}|$~\cite{BLL}.  Unlike the lepton energy
and hadronic invariant mass spectra, nonperturbative effects are only important
in the resonance region, and play a parametrically suppressed role when ${\rm
d}\Gamma / {\rm d}q^2$ is integrated over $q^2 > (m_B-m_D)^2$, which is
required to eliminate the $\bar B\to X_c \ell\bar\nu$ background.  We discuss
these backgrounds for $q^2$ slightly below $(m_B-m_D)^2$, and point out that
instead of $q^2 > (m_B-m_D)^2 = 11.6\,{\rm GeV}^2$, the cut can be lowered to
$q^2 \gtrsim 10.5\,{\rm GeV}^2$.  This is important experimentally,
particularly when effects of a finite neutrino reconstruction resolution are
included.
\end{Abstract}
\vfill
\begin{Presented}
5th International Symposium on Radiative Corrections \\ 
(RADCOR--2000) \\[4pt]
Carmel CA, USA, 11--15 September, 2000
\end{Presented}
\vfill
\end{titlepage}
\def\thefootnote{\arabic{footnote}}
\setcounter{footnote}{0}
A precise and model independent determination of the
Cabibbo-Kobayashi-Maskawa (CKM) matrix element $V_{ub}$ is important for
testing the Standard Model at $B$ factories via the comparison of the
angles and the sides of the unitarity triangle.  

If it were not for the huge background from decays to charm, it would be
straightforward to determine $|V_{ub}|$ from inclusive semileptonic
decays.  Inclusive $B$ decay rates can be computed model independently in
a series in $\lqcd/m_b$ and $\alpha_s(m_b)$ using an operator product
expansion (OPE)~\cite{CGG,incl,MaWi,Blok}, and the result may
schematically be written as
\begin{equation}\label{schematic}
{\rm d}\Gamma = \pmatrix{ b{\rm ~quark} \cr {\rm decay}\cr } \times 
  \bigg\{ 1 + \frac0{m_b} + \frac{f(\lambda_1,\lambda_2)}{m_b^2} + \ldots
  + \frac{\alpha_s}\pi\, (\ldots) + \frac{\alpha_s^2}{\pi^2}\, (\ldots) 
  + \ldots \bigg\} \,.
\end{equation}
At leading order, the $B$ meson decay rate is equal to the $b$ quark decay
rate.  The leading nonperturbative corrections of order $\lqcd^2 / m_b^2$
are characterized by two heavy quark effective theory (HQET) matrix
elements, usually called $\lambda_1$ and $\lambda_2$.  These matrix
elements also occur in the expansion of the $B$ and $B^*$ masses in powers
of $\lqcd/m_b$,
\begin{equation}\label{massrelation}
m_B = m_b + \bar\Lambda 
  - {\lambda_1 + 3 \lambda_2 \over 2m_b} + \ldots \,, \qquad
m_{B^*} = m_b + \bar\Lambda 
  - {\lambda_1 - \lambda_2 \over 2m_b} + \ldots \,.
\end{equation}
Similar formulae hold for the $D$ and $D^*$ masses.  The parameters
$\bar\Lambda$ and $\lambda_1$ are independent of the heavy $b$ quark mass,
while there is a weak logarithmic scale dependence in $\lambda_2$.  The
measured $B^*-B$ mass splitting fixes $\lambda_2(m_b) = 0.12\,{\rm
GeV}^2$, while $\bar\Lambda$ and $\lambda_1$ (or, equivalently, a short
distance $b$ quark mass and $\lambda_1$) may be determined from other
physical quantities \cite{FLS,gremmetal,bsg}.  Thus, a measurement of the
total $B\rightarrow X_u \ell\bar\nu$ rate would provide a $\sim 5\%$
determination of $|V_{ub}|$ \cite{upsexp,burels}.

Unfortunately, the $\bar B\to X_u\ell\bar\nu$ rate can only be measured
imposing cuts on the phase space to eliminate the $\sim 100$ times larger
$\bar B\to X_c\ell \bar\nu$ background.  Since the predictions of the OPE
are only model independent for {\it sufficiently inclusive} observables,
these cuts can destroy the convergence of the expansion.  This is the case
for two kinematic regions  for which the charm background is absent and
which have received much attention: the large lepton energy region, 
$E_\ell > (m_B^2-m_D^2)/2m_B$, and the small hadronic invariant mass
region, $m_X < m_D$ \cite{BKP,FLW,DU}. 

The poor behaviour of the OPE for these quantities is slightly subtle,
because in  both cases there is sufficient phase space for many different
resonances to be produced in the final state, so an inclusive description
of the decays is still appropriate. However, in both of these regions of
phase space the $\bar B\to X_u\ell\bar\nu$ decay  products are dominated
by high energy, low invariant mass hadronic states,
\begin{equation}\label{shapefnregime}
E_X\sim m_b,\ m_X^2\sim \lqcd m_b\gg\lqcd^2
\end{equation}
(where $E_X$ and $m_X$ are the energy and invariant mass of the final hadronic state).
In this region the differential rate is very sensitive to the details of
the wave function of  the $b$ quark in the $B$ meson.  Since the OPE is
just sensitive to local matrix elements corresponding to expectation
values of operators in the meson, the first few orders in the OPE do not
contain enough information to describe the decay, and as a result the OPE
does not converge.

This is simple to see by considering the kinematics.  A $b$ quark in a $B$
meson has momentum
\begin{equation}
p_b^\mu=m_b v^\mu+k^\mu
\end{equation}
where $v^\mu$ is the four-velocity of the quark, and $k^\mu$ is a small
residual momentum of order $\lqcd$.  If the hadron decays to leptons with
momentum $q$ and light hadrons with total momentum $p_X$, the invariant
mass of the light hadrons may  be written
\begin{equation}\label{kinexp}
m_X^2=(m_b v+k-q)^2=(m_b v-q)^2+2k\cdot(m_b v-q)+O(\lqcd^2).
\end{equation}
The first term in the expansion is $O(m_b^2)$ over most of phase space,
while the second is $O(\lqcd m_b)$, and so is suppressed over most of
phase space.  The OPE presumes that this power counting holds, so that the
second term may be treated as a small perturbation.  However, if $E_X$ is
large and $m_X$ is small, $m_b v-q$ is almost light-like,
\begin{equation}
m_b v^\mu-q^\mu=(E_X,0,0,E_X)+O(\lqcd)
\end{equation}
in the $b$ rest frame where $v^\mu=(1,0,0,0)$.  Since $E_X\sim O(m_b)$,
$(m_b v-q)^2=O(\lqcd m_b)$.  Thus, in this region the first two terms in
(\ref{kinexp}) are of the same order (but still parametrically larger than
the remaining terms), and the invariant mass of the final hadronic state
reflects the distribution of the light-cone component of the residual
momentum of the heavy quark in the hadron,
\begin{equation}
m_X^2=(m_b v-q)^2+2 E_X k_++\dots,\ k_+\equiv k_0+k_3 .
\end{equation}
Since the differential rate in this region depends on the invariant mass
of the final state, it is therefore sensitive at leading order to the
light-cone wave function of the heavy quark in the meson, $f(k_+)$.

In terms of the OPE, this light-cone wave function arises because of
subleading terms in the OPE proportional to $E_X\lqcd/m_X^2$, which are
suppressed over most of phase space but are $O(1)$ in the region
(\ref{shapefnregime}). It has been shown that the most singular terms in
the OPE may be resummed into a nonlocal operator whose matrix element in a
$B$ meson is the light-cone structure function of the meson.  Since
$f(k_+)$ is a nonperturbative function, it cannot be calculated 
analytically, so the rate in the region (\ref{shapefnregime}) is
model-dependent even at leading order in $\lqcd/m_b$.

\begin{figure}[tb]
\centerline{\epsfxsize=15truecm \epsfbox{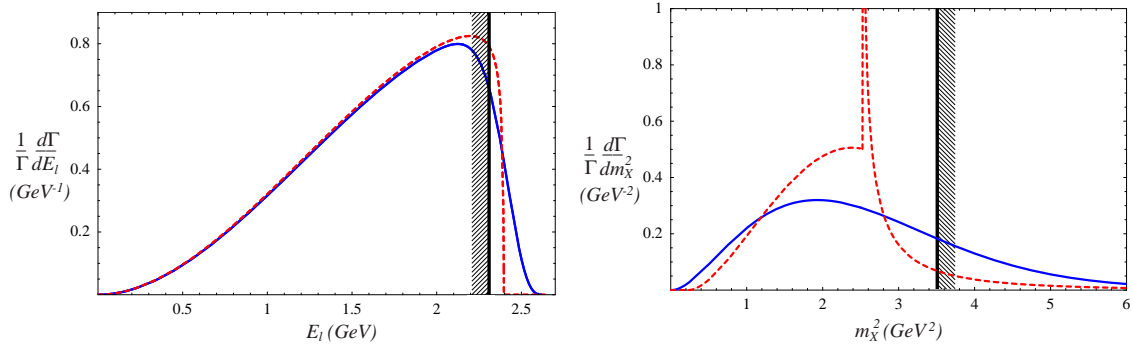} }
\vspace*{-.8cm}
\caption[]{The shapes of the lepton energy and hadronic invariant mass spectra.
The dashed curves are the $b$ quark decay results to ${\cal O}(\alpha_s)$,
while the solid curves are obtained by smearing with the model distribution
function $f(k_+)$ in Eq.~(\ref{sfn}).  The unshaded side of the vertical lines
indicate the region free from charm background.}
\label{twospectra}
\end{figure}

The situation is illustrated in Fig.~\ref{twospectra}, 
where we have plotted the lepton energy and hadronic invariant mass
spectra in the parton model (dashed curves) and incorporating a simple
one-parameter model for the distribution function (solid curves)~\cite{MN}
\begin{equation}\label{sfn}
f(k_+) = {32\over \pi^2 \Lambda}\, (1-x)^2\, 
  \exp\left[-{4\over \pi}(1-x)^2\right] \Theta(1-x) \,, 
\qquad  x \equiv {k_+\over \Lambda} \,, \qquad \Lambda=0.48\,{\rm GeV}\,.
\end{equation}
The differences between the curves in the regions of interest indicate the
sensitivity of the spectrum to the precise form of $f(k_+).$  Currently,
there are measurements of $|V_{ub}|$ from both methods.  From the lepton
energy cut, the PDG reports $|V_{ub}/V_{cb}|=0.08\pm 0.02$, while a recent
DELPHI measurement using the hadronic invariant mass cut gives
$|V_{ub}/V_{cb}|=0.103^{+0.011}_{-0.012}\,\mbox{(syst.)}\pm 0.016 
\,\mbox{(stat.)}\pm0.010\,\mbox{(theory)}$ \cite{delphi00}.  In both
cases, the theoretical error is an estimate based on varying different
models of $f(k_+)$, and so these measurements are no more
model-independent  than the exclusive measurement from
$B\rightarrow\rho\ell\bar\nu$.  While it may be possible in the future to
extract $f(k_+)$ from the $B\to X_s\gamma$ photon
spectrum~\cite{shape,LLR}, unknown order $\lqcd/m_b$ corrections arise
when relating this to semileptonic $b\rightarrow u$ decay, limiting the
accuracy with which $|V_{ub}|$ may be obtained.

\begin{figure}[tb]
\centerline{\epsfxsize=9truecm \epsfbox{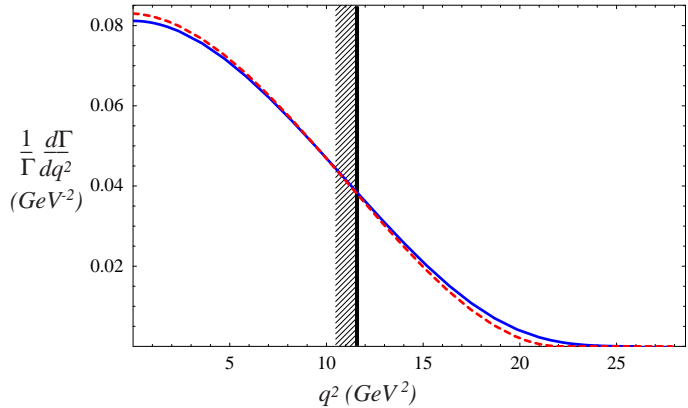} }
\vspace*{-.5cm}
\caption[]{The dilepton invariant mass spectrum.  
The notation is the same as in Fig.~\ref{twospectra}.}
\label{qsqspectrum}
\end{figure}

\begin{figure}[bt]
\centerline{\epsfxsize=15truecm \epsfbox{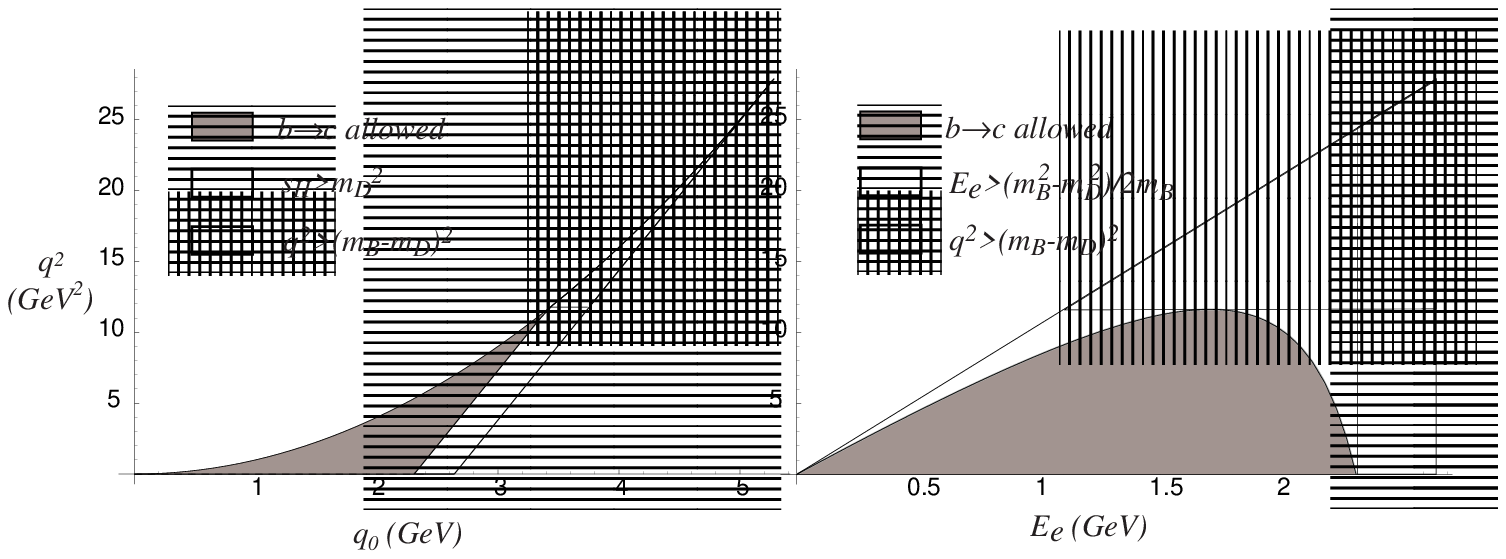} }
\vspace*{-.8cm}
\caption[]{Dalitz plots relevant for $\bar B\rightarrow
X_u\ell\bar\nu_\ell$.  The shaded regions indicate the part of phase
space where $\bar B\rightarrow X_c\ell\bar\nu_\ell$ background is
present, and the vertical dashed regions corresponds to the cut
$q^2>(m_B-m_D)^2$.  In the $q^2-q_0$ plane, the horizontal dashed
region corresponds to an invariant mass cut $m_X^2>m_D^2$, whereas
in the $q^2-E_\ell$ plane the horizontal dashed region corresponds to
the charged lepton energy cut $E_\ell>(m_B^2-m_D^2)/2 m_B$.  Note that
at tree level, $b\rightarrow u$ semileptonic decay populates the entire
triangle on the right-hand plot, but only the right boundary of the
left-hand plot.}
\label{dalitz}
\end{figure}

Clearly, one would like to be able to find a cut which eliminates the
charm background but does not destroy the convergence of the OPE, so that
the distribution function $f(k_+)$ is not required.  In Ref.~\cite{BLL} we
pointed out that this is the situation for a cut on the dilepton invariant
mass.  Decays with $q^2 \equiv (p_\ell + p_{\bar\nu})^2 > (m_B - m_D)^2$
must arise from $b\to u$ transition.  Such a cut forbids the hadronic
final state from moving fast in the $B$ rest frame, and simultaneously
imposes $m_X < m_D$ and $E_X < m_D$.  Thus, the light-cone expansion which
gives rise to the shape function is not relevant in this region of phase
space~\cite{DU,BI}.   The effect of smearing the $q^2$ spectrum with the
model distribution function in Eq.~(\ref{sfn}) is illustrated in
Fig.~\ref{qsqspectrum}.
It is clearly a subleading effect.  The Dalitz plots relevant for the
charged lepton energy and hadronic invariant mass cuts are shown in Fig.\
\ref{dalitz}.  Note that the region selected by a $q^2$ cut is entirely
contained within the $m_X^2$ cut, but because the dangerous region of high
energy, low invariant mass final states is not included with the $q^2$
cut, the OPE does not break down.  It is also important to note, however,
that the $q^2$ cut does make the OPE worse than for the full rate; as we
will show, the relative size of the unknown $\lqcd^3/m_b^3$ terms grows as
the $q^2$ cut is raised.  Equivalently, as was stressed in
\cite{Matthias},  the effective expansion parameter for this region is
$\lqcd/m_c$, not $\lqcd/m_b$.

The $\bar B\to X_u\ell\bar\nu$ decay rate with lepton invariant mass above
a given cutoff can therefore be reliably computed working to a fixed order
in the OPE (i.e., ignoring the light-cone distribution function),
\begin{eqnarray}\label{q2spec}
{1\over \Gamma_0}\, {{\rm d}\Gamma \over {\rm d}\qsh} &=&
  \bigg( 1 + {\lambda_1\over 2m_b^2} \bigg)\, 2\, (1-\qsh)^2\, (1+2\qsh) 
  + {\lambda_2\over m_b^2}\, (3 - 45\hat q^4 + 30\hat q^6) \nonumber\\*
&&{} + {\alpha_s(m_b) \over \pi}\, X(\qsh)
  + \bigg( {\alpha_s(m_b) \over \pi} \bigg)^2\, \beta_0\, Y(\qsh) + \ldots \,,
\end{eqnarray}
where $\qsh = q^2/m_b^2$, $\beta_0 = 11 - 2n_f/3$, and $\Gamma_0 = G_F^2\,
|V_{ub}|^2\, m_b^5 / (192\, \pi^3)$ is the tree level $b\to u$ decay
rate.  The ellipses in Eq.~(\ref{q2spec}) denote terms of order
$(\lqcd/m_b)^3$ and order $\alpha_s^2$ terms not enhanced by $\beta_0$. 
The function $X(\qsh)$ is known analytically~\cite{JK}, whereas $Y(\qsh)$
was computed numerically~\cite{LSW}.  The order $1/m_b^3$ nonperturbative
corrections are also known~\cite{m3corr}, as are the leading logarithmic
perturbative corrections proportional to $\alpha_s^n \log^n
(m_c/m_b)$~\cite{Matthias}.  The matrix element of the kinetic energy
operator, $\lambda_1$, only enters the $\hat q^2$ spectrum in a very
simple form, because the unit operator and the kinetic energy operator are
related by reparameterization invariance~\cite{LM}.

\begin{figure}[t]
\centerline{\epsfysize=5.4truecm \epsfbox{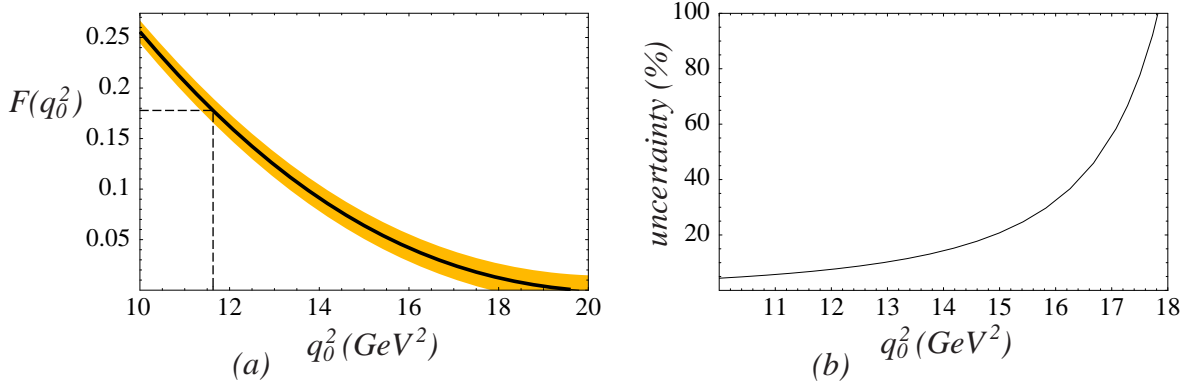} }
\vspace*{-.5cm}
\caption[]{(a) The fraction of $\bar B\to X_u \ell\bar\nu$ events with $q^2 >
\qcut^2$, $F(\qcut^2)$, in the upsilon expansion.  The dashed line
indicates  the lower cut $\qcut^2 = (m_B-m_D)^2 \simeq 11.6\,{\rm GeV}^2$,
which  corresponds to $F = 0.178 \pm 0.012$.  The shaded region is the
estimated  uncertainty due to $\lqcd^3 / m_b^3$ terms; which is shown in
(b) as a  percentage of $F(\qcut^2)$. }
\label{fractionplot}
\end{figure}

The relation between the total $\bar B\to X_u \ell\bar\nu$ decay rate and
$|V_{ub}|$ is known at the $\sim5\%$ level~\cite{upsexp,burels}, 
\begin{equation}\label{Vubups}
|V_{ub}| = (3.04 \pm 0.06 \pm 0.08) \times 10^{-3}\,
  \left( {{\cal B}(\bar B\to X_u \ell\bar\nu)|_{q^2 > \qcut^2} \over 
  0.001 \times F(\qcut^2) }\, {1.6\,{\rm ps}\over\tau_B} \right)^{1/2} \,,
\end{equation}
where $F(\qcut^2)$ is the fraction of $\bar B\to X_u \ell\bar\nu$ events
with $q^2 > \qcut^2$, satisfying $F(0)=1$.  The errors explicitly shown in
Eq.~(\ref{Vubups}) are the estimates of the perturbative and
nonperturbative uncertainties in the upsilon expansion~\cite{upsexp}
respectively.   At the present time the biggest uncertainty is due to the
error of a short distance $b$ quark mass, whichever way it is defined
\cite{Matthias}.  (This can be cast into an  uncertainty in an
appropriately defined $\bar\Lambda$, or the nonperturbative contribution
to the $\Upsilon(1S)$ mass, etc.)  By the time the $q^2$ spectrum in $\bar
B\to X_u \ell\bar\nu$ is measured, this uncertainty should be reduced from
extracting $m_b$ from the hadron mass~\cite{FLS} or lepton
energy~\cite{gremmetal} spectra in $\bar B\to X_c \ell\bar\nu$, or from
the photon energy spectrum~\cite{bsg} in $B\to X_s\gamma$.  The
uncertainty in the perturbation theory calculation will be largely reduced
by computing the full order $\alpha_s^2$ correction in
Eq.~(\ref{Vubups}).   The largest ``irreducible" uncertainty is from order
$\lqcd^3/m_b^3$ terms in the OPE, the  estimated size of which is shown in
Fig.~\ref{fractionplot}, together with our central value for $F(\qcut^2)$,
as functions of $\qcut^2$.

\begin{figure}[tb]
\centerline{\epsfysize=5truecm \epsfbox{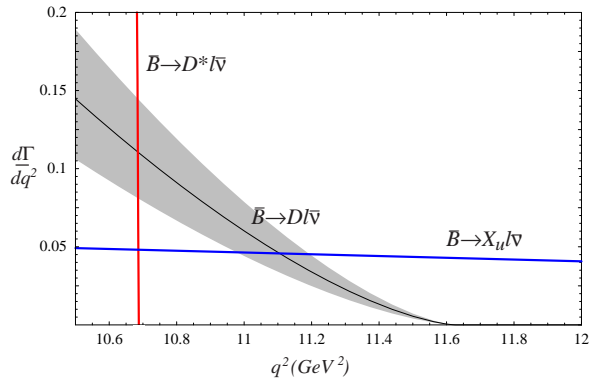} }
\vspace*{-.2cm}
\caption[]{Charm backgrounds near $q^2=(m_B-m_D)^2$ (arbitrary units). The
shaded region denotes the uncertainty on the $\bar B\rightarrow D\ell\bar\nu$ rate.}
\label{charmplot}
\end{figure}

There is another advantage of the $q^2$ spectrum over the $m_X$ spectrum
to measure $|V_{ub}|$.  In the variable $m_X$, about 20\% of the charm
background is located right next to the $b\to u$ ``signal region", $m_X <
m_D$, namely $\bar B\to D \ell\bar\nu$ at $m_X = m_D$.  In the variable
$q^2$, the charm background just below $q^2 = (m_B-m_D)^2$ comes from the
lowest mass $X_c$ states.  Their $q^2$ distributions are well understood
based on heavy quark symmetry~\cite{HQS}, since this region corresponds to
near zero recoil.  Fig.~\ref{charmplot} shows the $\bar B\to D \ell
\bar\nu$ and $\bar B\to D^* \ell \bar\nu$ decay rates using the measured
form factors~\cite{data} (and $|V_{ub}| = 0.0035$).  The $\bar B\to X_u
\ell\bar\nu$ rate is the flat curve.  Integrated over the region $q^2 >
(m_B-m_{D^*})^2 \simeq 10.7\, {\rm GeV}^2$, the uncertainty of the $B\to
D$ background is small due to its $(w^2-1)^{3/2}$ suppression compared to
the $\bar B\to X_u\ell\bar\nu$ signal.  This uncertainty will be further
reduced in the near future.  This increases the $b\to u$ region relevant
for measuring $|V_{ub}|$ by $\sim1\, {\rm GeV}^2$.  The $B\to D^*$ rate is
only suppressed by $(w^2-1)^{1/2}$ near zero recoil, and therefore it is
more difficult to subtract it reliably from the $b\to u$ signal.  The
nonresonant $D\pi$ final state contributes in the same region as $\bar
B\to D^*$, and it is reliably predicted to be small near maximal $q^2$
(zero recoil) based on chiral perturbation theory~\cite{Dpi}.  The
$D^{**}$ states only contribute for $q^2 < 9\, {\rm GeV}^2$, and some
aspects of their $q^2$ spectra are also known model
independently~\cite{Dss}.

Concerning experimental considerations, measuring the $q^2$ spectrum
requires reconstruction of the neutrino four-momentum, just like measuring
the hadronic invariant mass spectrum.  A lepton energy cut may be required
for this technique, however, the constraint $q^2 > (m_B-m_D)^2$
automatically implies $E_\ell > (m_B-m_D)^2/2m_B \simeq 1.1\,$GeV in the
$B$ rest frame.  Even if the $E_\ell$ cut has to be slightly larger than
this, the utility of our method will not be affected, but a calculation
including the effects of arbitrary $E_\ell$ and $q^2$ cuts would be
required.  If experimental resolution on the reconstruction of the
neutrino momentum necessitates a significantly larger cut than $\qcut^2 =
(m_B-m_D)^2$, then the uncertainties in the OPE calculation of
$F(\qcut^2)$ increase.  In this case, it may be possible to obtain useful
model independent information on the $q^2$ spectrum in the region $q^2 >
m_{\psi(2S)}^2 \simeq 13.6\,{\rm GeV}^2$ from the $q^2$ spectrum in the
rare decay $\bar B \to X_s \ell^+\ell^-$, which may be measured in the
upcoming Tevatron Run-II.  

In conclusion, we have shown that the $q^2$ spectrum in inclusive
semileptonic $\bar B \to X_u \ell \bar\nu$ decay gives a model independent
determination of $|V_{ub}|$ with small theoretical uncertainty. 
Nonperturbative effects are only important in the resonance region, and
play a parametrically suppressed role when ${\rm d}\Gamma/{\rm d}q^2$ is
integrated over $q^2>(m_B-m_D)^2$, which is required to eliminate the
charm background.  This is a qualitatively better situation than other
extractions of $|V_{ub}|$ from inclusive charmless semileptonic $B$ decay.

\section*{Acknowledgements}

This work was supported in part by the Natural Sciences and Engineering
Research Council of Canada and by the Director, Office of Science, 
Office of High Energy and Nuclear Physics, Division of High Energy Physics, 
of the U.S.\ Department of Energy under Contract DE-AC03-76SF00098.%

\end{document}